\documentclass[a4paper,11pt]{article}
\usepackage{pos}
\usepackage{subcaption}
\usepackage{lineno}
\title{Deep-learning-driven event reconstruction applied to simulated data from a single Large-Sized Telescope of CTA}
 \ShortTitle{Deep-learning-driven event reconstruction applied to a single LST}

\author*[a]{P. Grespan}
\author[c,d]{M. Jacquemont}
\author[a,b]{R. López-Coto}
\author[e]{T. Miener}
\author[e]{D. Nieto-Castaño}
\author[c]{T. Vuillaume}

\affiliation[a]{Istituto Nazionale di Fisica Nucleare, Sezione di Padova, I-35131, Padova, Italy}
\affiliation[b]{now at Instituto de Astrofísica de Andalucía, CSIC, 18080 Granada, Spain}
\affiliation[c]{Laboratoire
d’Annecy de Physique des Particules, Univ. Grenoble Alpes, Univ. Savoie Mont Blanc, CNRS, LAPP, 9 chemin de bellevue, 74940 Annecy, France}
\affiliation[d]{LISTIC, Université Savoie Mont-Blanc Polytech Annecy-Chambéry, 5 chemin de bellevue, 74940 Annecy, France}
\affiliation[e]{Instituto de Física de Partículas y del Cosmos and Departamento de EMFTEL, Universidad
Complutense de Madrid, Madrid, Spain}

\forColl{CTA LST Project} 

\emailAdd{pgrespan@pd.infn.it}
\emailAdd{pietro.grespan@studenti.unipd.it}

\abstract{When very-high-energy gamma rays interact high in the Earth’s atmosphere, they produce cascades of particles that induce flashes of Cherenkov light. Imaging Atmospheric Cherenkov Telescopes (IACTs) detect these flashes and convert them into shower images that can be analyzed to extract the properties of the primary gamma ray. The dominant background for IACTs is comprised of air shower images produced by cosmic hadrons, with typical noise-to-signal ratios of several orders of magnitude. The standard technique adopted to differentiate between images initiated by gamma rays and those initiated by hadrons is based on classical machine learning algorithms, such as Random Forests, that operate on a set of handcrafted parameters extracted from the images. Likewise, the inference of the energy and the arrival direction of the primary gamma ray is performed using those parameters. State-of-the-art deep learning techniques based on convolutional neural networks (CNNs) have the potential to  enhance the event reconstruction performance, since they are able to autonomously extract features from raw images, exploiting the pixel-wise information washed out during the parametrization process. 
Here we present the results obtained by applying deep learning techniques to the reconstruction of Monte Carlo simulated events from a single, next-generation IACT, the Large-Sized Telescope (LST) of the Cherenkov Telescope Array (CTA). We use CNNs to separate the gamma-ray-induced events from hadronic events and to reconstruct the properties of the former, comparing their performance to the standard reconstruction technique. Three independent implementations of CNN-based event reconstruction models have been utilized in this work, producing consistent results.}

\FullConference{37$^{\rm{th}}$ International Cosmic Ray Conference (ICRC 2021)\\
		July 12th -- 23rd, 2021\\
		Online -- Berlin, Germany}

\begin{document}
\maketitle

\section{Introduction}
When a very-high-energy (VHE, $20 $ GeV $ < E < 300$ TeV) photon reaches the Earth, its interaction with the air molecules generates a cascade of relativistic particles that produces a dim flash of Cherenkov light.
If an Imaging Atmospheric Cherenkov Telescope (IACT) is inside the Cherenkov light-pool of the gamma ray, the Cherenkov photons are collected by its reflector and focused onto a camera made e.g. of photomultipliers (PMTs) in such a way that their arrival directions are transformed in points of a (pixelized) shower image. Those images can then be analyzed to extract the properties of the primary gamma ray.

The flux of incoming gamma rays is dwarfed by the flux of charged cosmic rays (CRs) entering the atmosphere - mostly protons and Helium nuclei - that can mimic the signal which we are interested in, resulting in a signal-to-noise ratio of about 1/1000 for bright VHE gamma-ray sources. Fortunately, the different development of the cascades result in images of different shapes 
and can be discriminated.

The Cherenkov Telescope Array (CTA) is the observatory that will host the next generation of IACTs \cite{acharya}. It will be built in two sites, each hosting tens of telescopes: one will be located in the Northern hemisphere, at La Palma, the other one in the Southern hemisphere, in the Atacama desert,
in order to achieve full sky coverage.
Three types of IACTs will be deployed: the \textit{Large-Sized Telescope} (LST), the \textit{Medium-Sized Telescope} (MST) and the \textit{Small-Sized Telescope} (SST). The LST, with its 23m diameter detector and field of view of 4.3 degree diameter, is optimised to reach the lowest achievable energy thresholds. The LST prototype is currently taking commissioning data at La Palma.

In these proceedings we will investigate the application of state-of-the-art deep learning techniques to the analysis of simulated images recorded by a single LST.


\section{Event reconstruction for LST}
In the context of IACTs, the event reconstruction consists of successfully fulfilling three tasks: \textit{(i)} separate gamma-ray events from hadronic events, \textit{(ii)} reconstruct the energy of the gamma rays, \emph{(iii)} reconstruct their arrival direction.

\begin{figure}
    \centering
    \includegraphics[width=0.9\linewidth]{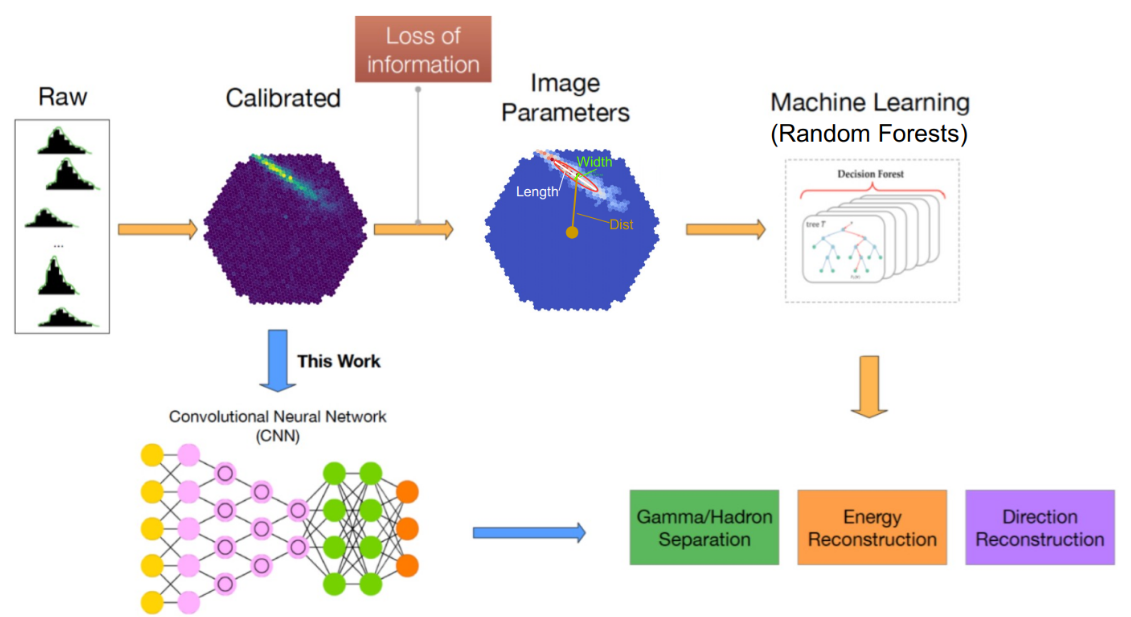}
    \caption{Standard LST event reconstruction scheme (orange arrows) compared to the deep learning approach. After the raw analog signal is calibrated and the image generated, cleaning and parameter extraction are applied to feed the Random Forests, while this step is skipped when using CNNs, accessing the pixel-wise information.}
    \label{fig:thiswork}
\end{figure}

The classical routine for the event reconstruction from images taken with LST, also illustrated in Fig. \ref{fig:thiswork}, is based on classical machine learning algorithms called Random Forests (RFs), similarly to the one adopted for the MAGIC telescopes (more detail about this analysis can be found in \cite{rfs}).
Once the telescope is triggered, the raw analog signal of each PMT is sampled by an analog to digital converter (ADC) and stored by the data acquisition system; from these raw data, signal is integrated and the total number of photons (charge) and their averaged arrival times (peak times) in the PMT are estimated, thus producing an image of two layers (charge and time).
After the image is produced, a \textit{cleaning procedure} is applied in order to get rid of the pixels dominated by the night sky background light and keep only those actually containing the shower image. Then, from the cleaned image a set of parameters containing relevant information is extracted (e.g. \texttt{intensity}, \texttt{width}, \texttt{length}, timing information, etc.).

Three Random Forest models - one per each of the aforementioned tasks - are trained on parameters extracted from simulated images. One is for event classification (gamma-like or hadronic), whose output, called \texttt{gammaness}, is a parameter that can be interpreted as the likelihood that the image belongs to a gamma-ray event. The other two are trained to predict \textit{energy} and \textit{arrival direction} of the gamma rays respectively. These models are then applied to the images recorded by the telescope.

The aim of this work is to perform and possibly improve this analysis skipping the cleaning and parametrization step and using convolutional neural networks, in order to exploit the maximum information contained in all the pixels to identify the signal. 

\subsection{Convolutional Neural Networks}
Differently from Random Forests, recent deep learning techniques are able to autonomously learn
how to extract information from raw data, deciding by themselves which patterns of the
dataset are meaningful for the task addressed. Convolutional neural networks (CNNs) \cite{lecun} are particular deep learning methods that show their full power in image recognition. They look for relevant features of the (raw) image by performing several convolutions with small square filters, and in the end convert the pixel information into the abstract information at the output layer. This output is a prediction of the quantity of interest (in our case \texttt{gammaness}, energy or arrival direction).


\section{Setup}
Three teams were involved in this project, performing three independent and different implementations of CNN-based event reconstruction models: PdVGG, CTLearn-TRN, and $\gamma$-PhysNet.
\\
\\
\textbf{PdVGG} \quad PdVGG, developed by INFN Padova, addresses the event reconstruction with a single-task approach, i.e. training one network for particle classification, one for energy and one for direction reconstruction.
 Since state-of-the-art API for CNNs take rank 3 tensors as input (rectangular images, square pixels), and the LST camera is hexagonal composed of hexagonal pixels, images need to be appropriately preprocessed in order to feed the networks.
To perform this step, images were fitted using bicubic interpolation and then resampled onto a rectangular lattice.
A handcrafted network of 13 layers 
based on the Visual Geometry Group architecture (VGG) \cite{vgg} has been implemented. VGG is a solid, popular architecture, often used as a baseline model in the deep learning community.
Among the different techniques used, the Cyclic Learning Rate technique \cite{clr} allows the network to explore different minima of the so-called loss function\footnote{A function that, during the training of a CNN, measures how much the predicted values deviate from the true, a-priori known values. Thus, the goal of the algorithm is to find the best parameters of the network that minimize this function.}, obtaining different network configurations in one single training.
\\
\\
\textbf{CTLearn-TRN} \quad Developed within the CTLearn framework\footnote{\href{https://github.com/ctlearn-project/ctlearn}{https://github.com/ctlearn-project/ctlearn}} \cite{ctlearn}, this implementation also uses single-task approach to tackle the event reconstruction.
The network is based on an architecture called Thin-ResNet (TRN) of 34 layers. ResNet \cite{resnet} is one of the most famous and powerful architectures for classification nowadays. An attention mechanism (or content aware mechanism) called squeeze and excitation \cite{squeeze} was also used: during the training it helps the network to put emphasis on the most relevant features of the image representation that is found at each layer.
The images were preprocessed with linear interpolation.
\\
\\
\textbf{$\gamma$-PhysNet} \quad
$\gamma$-PhysNet DA \cite{gammalearn}, developed using the GammaLearn framework\footnote{\href{https://gitlab.lapp.in2p3.fr/GammaLearn/GammaLearn}{https://gitlab.lapp.in2p3.fr/GammaLearn/GammaLearn}}, is a multi-task architecture, i.e. one single network able to perform the full event reconstruction, thus accomplishing all the three tasks of particle classification, energy and arrival direction reconstruction.
The networks are fed using IndexedConv \cite{indexedconv}, a technique that, knowing the nearest neighbor of each pixel, allows to perform convolution and pooling operations on non-Euclidean grid of data, with no need for interpolation.
The model is based on a ResNet architecture of 56 layers. Like CTLearn, it uses squeeze-and-excitation technique, plus another attention mechanism called Dual Attention (DA) and other techniques.
\subsection{Dataset}
The dataset used has been obtained with \texttt{CORSIKA v6.9} \cite{corsika} and its \texttt{IACT/ATMO} extension\footnote{\href{https://www.mpi-hd.mpg.de/hfm/~bernlohr/iact-atmo/}{https://www.mpi-hd.mpg.de/hfm/~bernlohr/iact-atmo/}} and the \texttt{sim-telarray v2018-11-07} package \cite{simtelarray}, then processed with \texttt{DL1 Data Handler v0.7.4}\footnote{\href{https://zenodo.org/record/4575505\#.YNNMI1kzaV5}{https://zenodo.org/record/4575505\#.YNNMI1kzaV5}} and \texttt{ctapipe v0.6.2}\footnote{\href{https://cta-observatory.github.io/ctapipe/}{https://cta-observatory.github.io/ctapipe/}}. It is composed of \emph{full simulated events} as recorded by a \textit{single} LST at LaPalma pointing at 20 deg. zenith angle in the South direction, and for air showers from primary protons, electrons, diffuse gamma-rays and pointlike gamma rays.
Diffuse events come from random directions within a view-cone of 10 degrees, while all pointlike gamma rays come from a single specific source in the sky at a 0.4 degree offset from the center of the camera. A set of $\sim 9 \cdot 10^5$ diffuse gamma rays and  $\sim 9 \cdot 10^5$ protons was used to train the network, while a set of  $\sim 6 \cdot 10^5$ electrons,  $\sim 2.5 \cdot 10^5$ protons and  $10^6$ pointlike gamma rays was used to test the performances.

\begin{figure}
\begin{subfigure}{.5\textwidth}
  \centering
  \includegraphics[width=1.\linewidth]{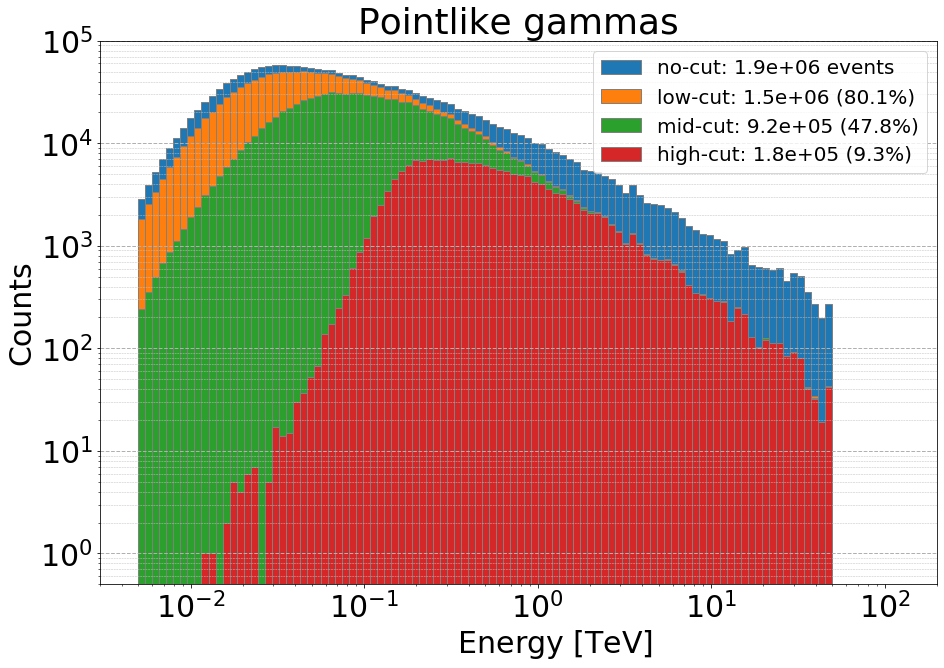}
  \caption{}
  \label{fig:orig_time}
\end{subfigure}%
\hfill
\begin{subfigure}{.5\textwidth}
  \centering
  \includegraphics[width=1.\linewidth]{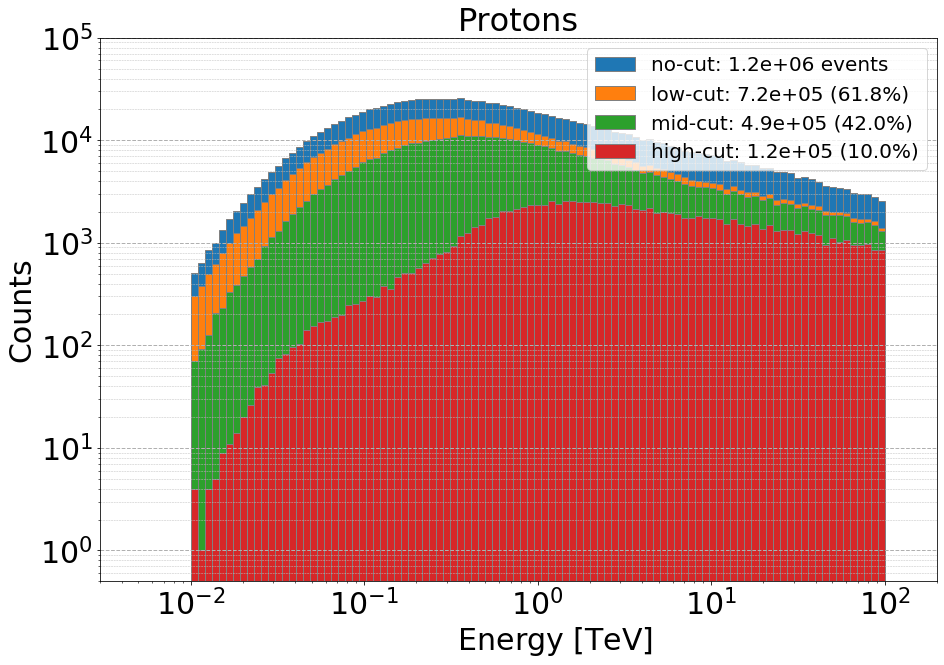}
  \caption{}
  \label{fig:interp_time}
\end{subfigure}
\caption{Spectra of pointlike gamma-ray (a) and proton (b) events (after trigger) as a function of the simulated energy, obtained using  Higher \texttt{intensity} cuts translate into higher energy thresholds.}
\label{fig:spectrum}
\end{figure}
\textbf{Data selection} \quad The dataset comprises many faint images, especially at lower energies, and many ($\sim 30 \%$) with shower image not well contained in the camera - thus selection cuts have been applied, depending on \textit{(i)} the \texttt{intensity} parameter, namely the total charge in photoelectrons (phe) contained in the image, and \textit{(ii)} the \texttt{leakage2} parameter, defined as the fraction of the image intensity contained in the two outermost rings of pixels. Four different image quality levels were defined on which the CNNs were trained and tested: \textbf{\texttt{no-cuts}} comprising all the images, \textbf{\texttt{low-cuts}} (\texttt{intensity} $> 50$ phe, \texttt{leakage2} $<0.2$), \textbf{\texttt{mid-cuts}} (\texttt{intensity} $> 200$ phe, \texttt{leakage2} $<0.2$) and \textbf{\texttt{high-cuts}} (\texttt{intensity} $> 1000$ phe, \texttt{leakage2} $<0.2$).
The effect of the different cuts on the dataset can be observed on the event energy distributions shown in Fig. \ref{fig:spectrum}: selecting higher-quality images comes at the price of discarding many low-energy events ($\sim 90 \%$ using \texttt{high-cuts}). Since intensity is roughly proportional to the energy of the gamma ray, each level is focused on a specific energy range - indeed, a higher \texttt{intensity} threshold translates into a higher energy threshold.

\section{Results}
Once the networks performed the event reconstruction on the test images, Instrument Response Functions (IRFs) have been computed using the \texttt{pyirf}\footnote{\href{https://github.com/cta-observatory/pyirf}{https://github.com/cta-observatory/pyirf}} library (a prototype for IRF generation for CTA), selecting gamma-like events by performing cuts on the \texttt{gammaness} parameter and on the $\theta^2$ parameter (that is, the squared angular separation between the simulated and reconstructed source position). These cuts are different for each energy bin, optimized to reach the best sensitivity. Sensitivity is defined as the minimum flux from a point-like source that it is able to detect with a 5-$\sigma$ significance, with additional constraints in the number of excess events and signal-to-noise ratio as defined in CTA performances\footnote{\href{https://www.cta-observatory.org/science/cta-performance/}{https://www.cta-observatory.org/science/cta-performance/}}. Random Forest models were produced and tested on the same dataset (using \texttt{low-}, \texttt{mid-} and \texttt{high-cuts}), in order to compare our results to the classical analysis, using \texttt{cta-lstchain v.0.5.2}\footnote{\href{https://github.com/cta-observatory/cta-lstchain}{https://github.com/cta-observatory/cta-lstchain}} and the source-independent analysis (see \cite{ruben} for further reference).

\begin{figure}
    \centering
    \includegraphics[width=\linewidth]{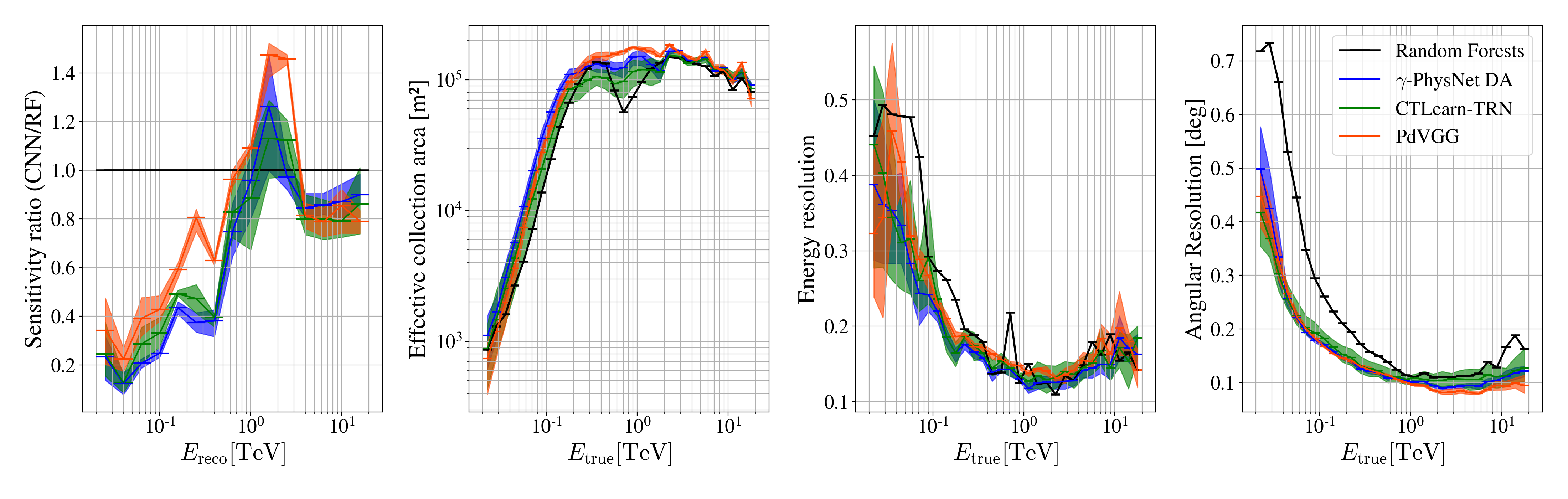}
    \caption{Instrument response functions obtained using \texttt{low-cuts} level. From left to right: ratio between CNN and RF sensitivities (lower is better); effective collection area (higher is better); energy resolution (lower is better); angular resolution (lower is better). Error bands are given by the $16^\text{th}$ and $84^\text{th}$ percentiles around the median of ten curves obtained training the CNNs ten times.}
    \label{fig:low_cut}
\end{figure}

\begin{figure}
    \centering
    \includegraphics[width=\linewidth]{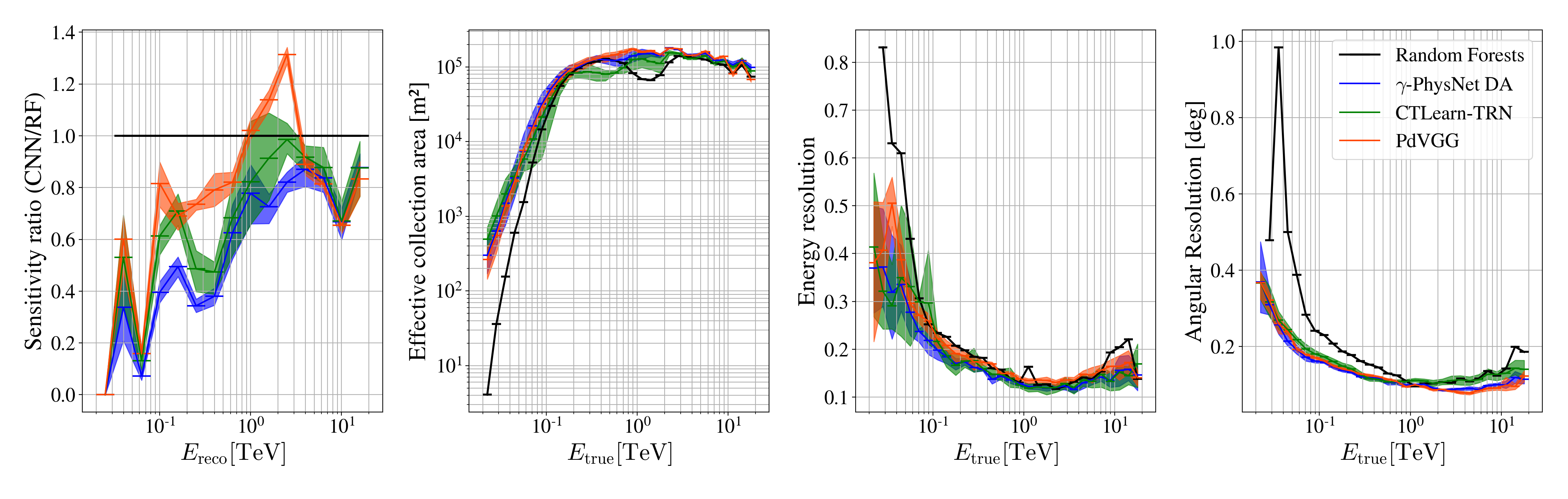}
    \caption{Instrument response functions obtained using \texttt{mid-cuts} level.}
    \label{fig:mid_cut}
\end{figure}

\begin{figure}
    \centering
    \includegraphics[width=\linewidth]{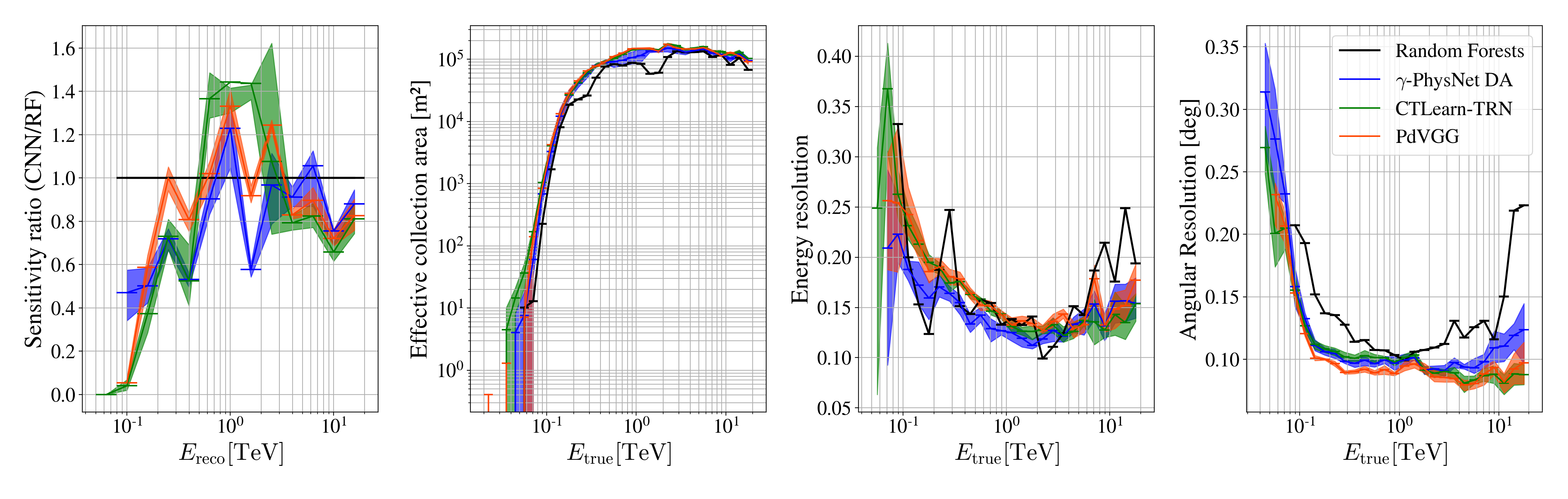}
    \caption{Instrument response functions obtained using \texttt{high-cuts} level.}
    \label{fig:high_cut}
\end{figure}

The figures of merit obtained using \texttt{low-cuts}, \texttt{mid-cuts} and \texttt{high-cuts} respectively are shown in Fig. \ref{fig:low_cut}, Fig. \ref{fig:mid_cut} and Fig. \ref{fig:high_cut}.
Despite some small difference, the performances obtained with the three implementations are very similar. Energy resolution is defined for each energy bin as the $68^\text{th}$ percentile of the distribution $|{E_\text{reco} - E_\text{true}}|/E_\text{true}$; even using different cuts, the curves show the same trend and similar performances across all the energy ranges, especially in the 500 GeV$-$10 TeV range where a resolution between $13\% - 18\%$ is reached, while for energies below 50 GeV it spans between $20\% - 40\%$. The angular resolution, computed as the square root of the $68^\textbf{th}$ percentile of the $\theta ^2$ distribution (that is the angular radius of the circle centered on the simulated source position containing $68\%$ of the reconstructed gamma-ray events), ranges between $0.3^\circ - 0.5^\circ$ at low energies below 50 GeV using \texttt{low-cuts}, down to a tenth of degree or less in the range above $2$ TeV. Angular resolution is sensitive
to the different quality cuts up to 1 TeV, while for higher energies the curves reach the same values.
The effective collection areas, defined as the geometrical area around the telescope in which a gamma ray triggers the instrument, are also very similar, spanning between $10^3 - 10^5$ m$^2$ at energies below 200 GeV, and flattening around $1.5 \cdot 10^5$ m$^2$ for higher energies (using \texttt{low} and \texttt{mid-cuts}; with \texttt{high-cuts}, the collection area is considerably smaller below 1 TeV, due to the high percentange of discarded events). 
In order to test the stability of the results, each team conducted ten times each training of the networks, using ten random seeds to initialize the parameters of the CNNs: the error bands displayed in the graphs are computed as the $16^\text{th}$ and $84^\text{th}$ percentiles around the median of the ten curves.

The performances are compatible with the ones obtained the Random Forests, although we can see a general enhancement brought by the CNNs using \texttt{low-} and \texttt{mid-cuts}, especially at the lower energies below 200 GeV. CNNs clearly perform better in the direction reconstruction task, where for example the Random Forests are overtaken up to more than 40\%.
The collection areas computed using the Random Forests and CNNs are similar, although the former exhibit a dip at energies around 700 GeV - 1 TeV. This is explained by more severe \texttt{gammaness} cuts in these bins: indeed, optimization for the best sensitivity led to different cuts for each model (CNN of RF), and the collection area is proportional to gamma efficiency (i.e. the ratio between the number of selected gamma candidates and the number of true simulated gamma-ray events for each bin). Likewise, more loose cuts in the mid-energy bins explain the larger collection area obtained with PdVGG.
Finally, CNNs show better differential sensitivity at energies below 1 TeV (which is the target range of LSTs) and above 3 TeV, while they tend to give worse performances in the $1-3$ TeV window.


\section{Conclusions}
In these proceedings we described three independent implementations of CNN-based event reconstruction models applied to the full event reconstruction of one single LST. We showed that all three lead to very consistent results, confirming that CNNs are an 
effective and reliable tool when applied to simulated data. We compared the results to those obtained with the standard RF technique, showing that CNNs are able to bring improvements to the performances, especially at low energies, probably thanks to the ability to exploit the pixel-wise information that is not accessible for Random Forests.

It is important to mark once again that these results were obtained using simulations, and we cannot infer that the performances will be the same when applied to data. In general, CNN analysis is less robust against variations in data due to small effects present in the data that are not properly simulated, as the noise present in each field of view the telescope is looking at. A first CNN-based analysis of real data from the LST-1 prototype is presented in another contribution of these proceedings \cite{thomas}.
Moreover, \texttt{cta-lstchain} standard analysis has been recently optimized (the current \texttt{v0.7} performs better than \texttt{0.5.2}), and also there is clear improvement when using the source-dependent analysis instead of the source-independent one used for this comparison (see \cite{ruben} for further reference).
Deep learning is a very dynamic and active research field, and there is surely room for further enhancement of CNN-based analyses, via testing new, powerful state-of-the-art deep learning techniques, or for example exploring custom models concatenating in the input both the images and the extracted parameters.

\subsection*{Acknowledgements}
This work was conducted in the context of the CTA Consortium. We gratefully acknowledge financial support from the agencies and organizations listed here:\\
\href{http://www.cta-observatory.org/consortium\_acknowledgments}{http://www.cta-observatory.org/consortium\_acknowledgments}.

\setlength{\bibsep}{0pt plus 0.3ex}

\clearpage
\section*{Full Authors List: \Coll\ }
\scriptsize
\noindent

H. Abe$^{1}$,
A. Aguasca$^{2}$,
I. Agudo$^{3}$,
L. A. Antonelli$^{4}$,
C. Aramo$^{5}$,
T.  Armstrong$^{6}$,
M.  Artero$^{7}$,
K. Asano$^{1}$,
H. Ashkar$^{8}$,
P. Aubert$^{9}$
% ...
\\

\noindent
$^{1}$Institute for Cosmic Ray Research, University of Tokyo.
$^{2}$Departament de Física Quàntica i Astrofísica, Institut de Ciències del Cosmos, Universitat de Barcelona, IEEC-UB.
$^{3}$Instituto de Astrofísica de Andalucía-CSIC.
$^{4}$INAF - Osservatorio Astronomico di Roma.
$^{5}$INFN Sezione di Napoli.
$^{6}$Aix Marseille Univ, CNRS/IN2P3, CPPM.
$^{7}$Institut de Fisica d'Altes Energies (IFAE), The Barcelona Institute of Science and Technology.
$^{8}$IRFU, CEA, Université Paris-Saclay.
$^{9}$LAPP, Univ. Grenoble Alpes, Univ. Savoie Mont Blanc, CNRS-IN2P3, Annecy.
% ...

%


\begin{thebibliography}{99}

{\footnotesize
\bibitem{acharya}
Acharya, B. et al. (2013). \textit{Introducing the CTA concept}. Astropart. Phys., 43, 3–18. 
\bibitem{rfs}
Albert, J. et al. (2008). \textit{Implementation of the random forest method for the
imaging atmospheric cherenkov telescope magic.} Nuclear Instruments and Methods in Physics Research Section A: Accelerators, Spectrometers, Detectors and Associated Equipment, 588(3), 424–432. \href{https://doi.org/10.1016/j.nima.2007.11.068}{https://doi.org/10.1016/j.nima.2007.11.068}
\bibitem{lecun}
LeCun, Y. et al. (1998). 
\textit{Gradient-based learning applied to document
recognition}. Proceedings of the IEEE, 86(11), 2278–2324.
\bibitem{ctlearn}
Nieto Castano, D. et al. (2019). \textit{CTLearn: Deep Learning for Gamma-ray Astronomy}. \href{https://arxiv.org/abs/1912.09877}{arxiv:1912.09877}
\bibitem{vgg}
Simonyan, K. et al. (2015). \textit{Very deep convolutional networks for large-scale image recognition.} \href{https://arXiv.org/abs/1409.1556}{arXiv:1409.1556}
\bibitem{clr}
Smith, L. et al. (2017). \textit{Cyclical learning rates for training neural networks}. \href{https://arxiv.org/abs/1506.01186}{arxiv.org:1506.01186}

\bibitem{resnet}
He, K. et al. (2015). \textit{Deep residual learning for image recognition.} \href{https://arXiv.org/abs/1512.03385}{arXiv:1512.03385}
\bibitem{squeeze}
Hu, J. et al. (2019). \textit{Squeeze-and-excitation networks.} \href{https://arXiv.org/abs/1709.01507}{arXiv:1709.01507}
\bibitem{gammalearn}
Jacquemont, M. et al. (2021). \textit{Multi-task architecture
with attention for imaging atmospheric cherenkov telescope data analysis,}. In Proceedings of the 16th International Joint Conference on Computer Vision, Imaging and Computer Graphics
Theory and Applications, INSTICC. SciTePress, 2021.
\bibitem{indexedconv}
Jacquemont, M. et al. (2019). \textit{Indexed operations for non-rectangular lattices applied to convolutional neural networks}. 
In Proceedings of the 14th International Joint Conference on Computer Vision, Imaging and
Computer Graphics Theory and Applications - Volume 5: VISAPP, INSTICC. SciTePress,
2019, pp. 362–371.
\bibitem[]{ruben}
Lopez-Coto, R. et al. \textit{Physics Performance of the Large-Sized Telescope prototype of the Cherenkov Telescope Array}. In these proceedings
\bibitem[]{corsika} 
Heck, D. et al. (1998). \textit{CORSIKA: a Monte Carlo code to simulate extensive air showers.}

\bibitem[]{simtelarray}
Bernlöhr, K. \textit{Simulation of imaging atmospheric Cherenkov telescopes with CORSIKA and sim\_telarray}. Astroparticle Physics 30 (2008) 149--158.

\bibitem{thomas}
Vuillaume, T. et al. \textit{Analysis of the Cherenkov Telescope Array first Large-Sized Telescope real data using convolutional neural networks}. In these proceedings
}








\end{thebibliography}
\end{document}